# Mysteries of Visual Experience

Jerome Feldman

'The most beautiful and profound experience is the feeling of mystery. It underlies religion as well as all deeper aspirations in art and science.' Einstein

## Introduction

*Update 2022*

*This arXiv paper has been in circulation for six years and has received modest, although not widespread attention. This revision changes very little of the original text; changes are shown in italics and are mainly focused updating the discussions and the references. Subsequent work is described in: On the Evolution of Subjective Experience* http://arxiv.org/abs/2008.08073 *, which is also revised. That paper also discusses recent advances in understanding subjective experience(SE) and applications in robotics, medicine and prosthetics. The (2016) first version of this article focused on the whole visual system and showed that the SE of a full detailed scene is* incompatible *with current cognitive neuroscience. However, it did not explicitly address the various levels of processing or the relation between foveal and peripheral vision*

An updated summary of all this work has been published as:
Feldman, J. (2022). Computation, perception, and mind. *Behavioral and Brain Sciences, 45*, E48. doi:10.1017/S0140525X21001886
A more readable, open access, version is: https://escholarship.org/uc/item/6cs78450

Science is a crowning glory of the human spirit and its applications remain our best hope for social progress. However, there are limitations to existing science and perhaps to any science. The general mind-body problem is known to be currently intractable and mysterious (8). This is one of many deep problems that are generally agreed to be beyond the present purview of Science, including many quantum phenomena, etc. However, all of these famous unsolved problems are either remote from everyday experience (entanglement, dark matter) or are hard to even define sharply (phenomenology, consciousness, etc.).

In this note, we consider some computational problems in vision that arise every time that we open our eyes and yet are demonstrably i*nconsistent* with current theories of neural computation. The focus will be on two famous related phenomena, known as the neural binding problem and the experience of a detailed stable visual world. I, among many others, have struggled with these issues for more than fifty years (1, 2, 3). Somewhat paradoxically, the continuing progress in scientific methods and knowledge reveals that these are both unsolvable within existing neuroscience. By considering some basic facts about how the brain processes image input, we will show that, under the standard theory, there are not nearly enough brain neurons to compute what we experience as vision. Inconsistencies like the ones shown here have had a profound effect on paradigm change in the sciences. More directly, the discussions below suggest possible new theories and experiments.

Demonstrations

The visual system can only capture fine detail in a small ( ~1 degree) part of the visual field; this is about the size of your thumbnail at arm’s length. “The experience of a detailed full-field stable visual world” refers to our subjective perception of a large high-resolution scene. First, consider Figure 1. Your vision is best in the center of gaze and the small letters in the center of the figure are easy to read when you look directly at them, but not when you look to the side. The letters further from the center are progressively larger and this describes how much coarser your vision becomes with eccentricity.

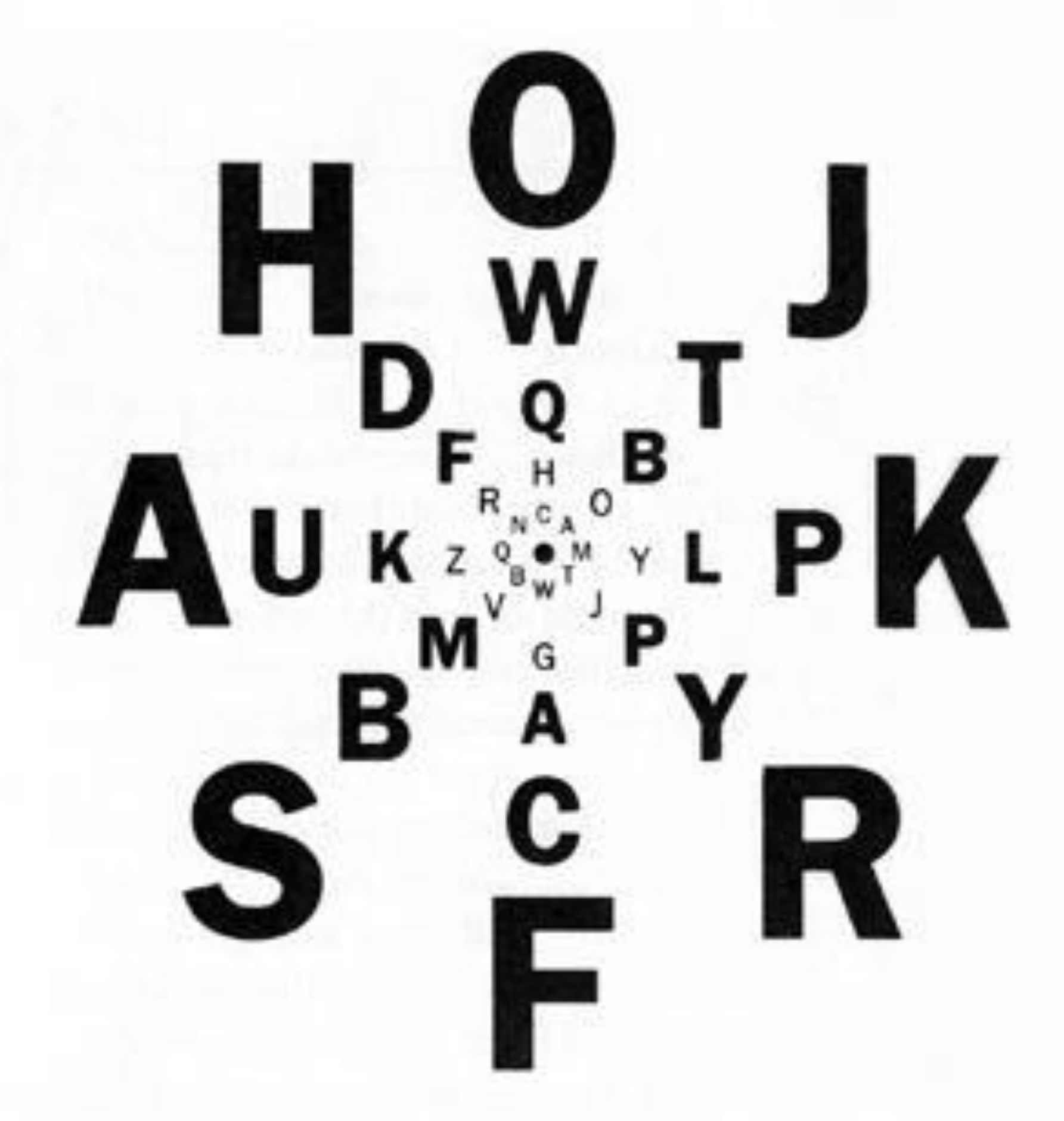

Figure 1. Size for Equal Visibility with Eccentricity

You can experience this directly using the line of text in Figure 2. Cover or close one eye and focus on the + in the center from a distance of about 12 inches. While holding focus, try to name the letters to the left. You should be able to do much better with the progressively larger letters to the right of the +. In ordinary viewing, there is no problem because we change our gaze several times a second.

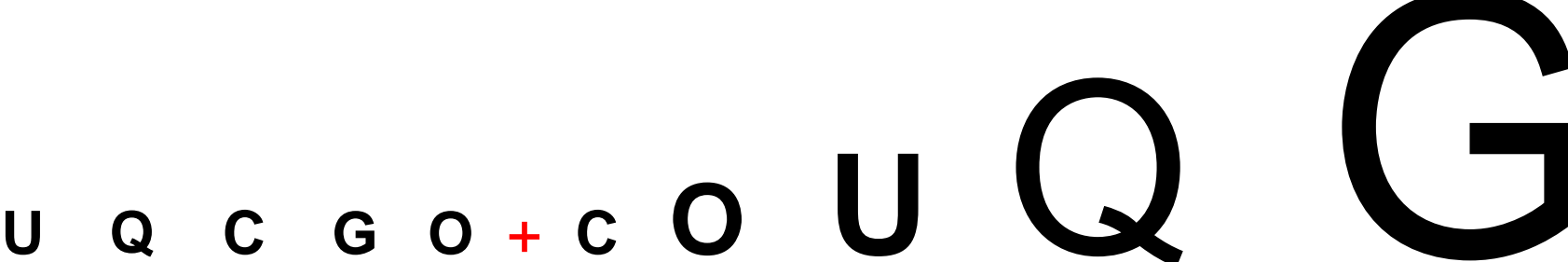

Figure 2. Demonstration of Visibility with Eccentricity

More generally, representing more information requires more hardware, which is why new phones are marketed as having cameras with more megapixels. This is also true for the neurons in the brain and this fact will play a major role in the discussion. There is a great deal known about how the brain processes visual information, largely because other mammals, particularly primates, have quite similar visual systems. We will focus on primary visual cortex or V1. Looking ahead, Figure 4B shows a flattened and projected view of the human brain with V1 on the far left.

Unsurprisingly, the brain realizes its high central resolution using many, densely packed, neurons. The central portion of the retina in the eye is the fovea and the downstream target of these foveal neurons in V1 of the brain is called the foveal projection.

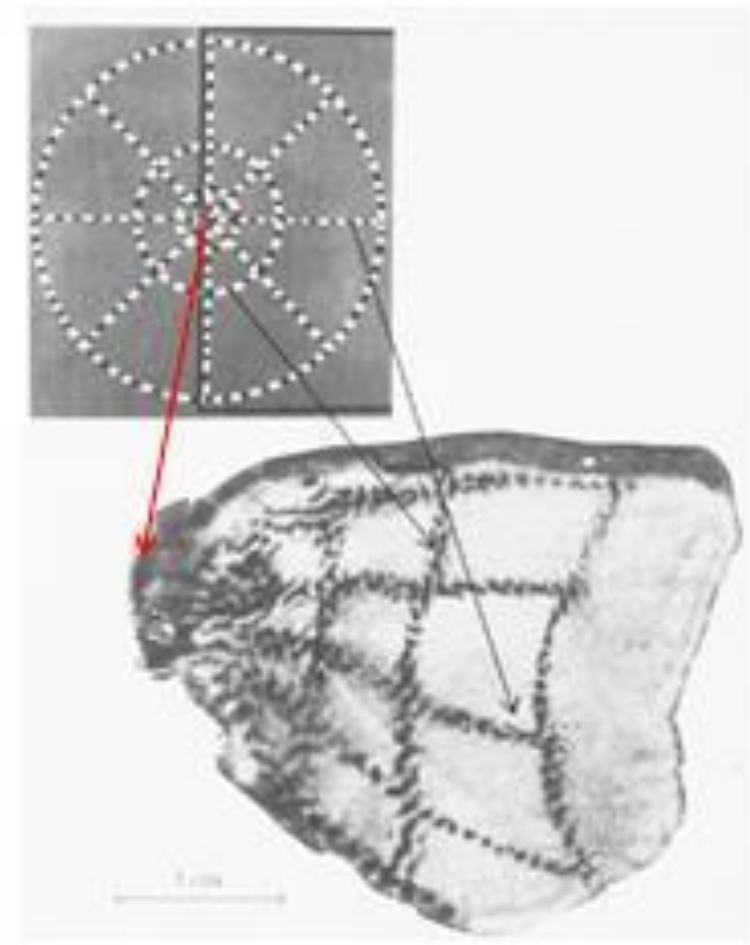

Figure 3. Tootell, et al. Experiment (4)

An important aspect of this architecture can be seen in Figure 3. The upper part of the figure depicts an oscillating radial stimulus, also with more detail in the center, which was presented to a primate subject. The lower half of the figure shows the parts of visual cortex that responded strongly to the input. As you can see, by far the most activity is the foveal projection on the far left, corresponding to the detailed image in the center of the input stimulus (red arrow). So, vision is most accurate in a small central area of the visual field and this is achieved by densely packed neurons in the corresponding areas of the brain (4).

However, our visual experience is not at all like this. We experience the world as fully detailed and there is currently no scientific explanation of this. There is more - we normally move (saccade) our gaze to new places in the scene about 3 or 4 times per second. These saccades help us see and act effectively and are not random. Again, our experience does not normally include any awareness of the saccades or the radically different visual inputs that they entail. Taken together these unknown links between brain and experience are known as “the experience of a detailed stable visual world” and this is generally accepted, if not understood.

There is extensive continuing research on various aspects of visual stability (6, 7, 8, 9, 10, 11). None of this work attempts to provide a complete solution and it is usually explicit that deep mysteries remain. Reference 9 is an excellent survey of behavioral findings and reference 11 has current neuroscience results.

We are attempting here to establish a much stronger statement. These stable world phenomena and a number of others are **inconsistent** with standard theories of neural processing (20). The demonstrations below involve combining findings from several distinct areas of investigation, as an instance of Unified Cognitive Science (12). Before digging in to the computational details, we consider some consequences of establishing that there is presently no scientific explanation of such visual experiences.

Why Inconsistency is important

It may seem surprising that we can exhibit such strong results without directly considering the relation between subjective (1st person) and objective (3rd person) experience, which is one of the core mysteries of the mind. For at least the cases presented, there is NO theory of the 1st person experience that is consistent with the known theory, structure, and behavior of the brain. This also challenges several proposed mind/brain relations as epiphenomenalism or the notion that the 1st/3rd person link is a brute fact of nature that cannot be further analyzed. These would not yield inconsistency.

In addition, throughout the history of science, crucial instances of inconsistency have led to profound reconsiderations and discoveries. One of the best-known cases is the fact that Rutherford's planetary model of the atom entailed that electrons rotating around the atomic nucleus would radiate energy and eventually crash into it. This was one of a number of deep inconsistencies leading to the development of quantum theory. A recent important inconsistency in Cognitive Science is the "Word Superiority Effect" (34). A wide range of experiments established that people were faster and more accurate in recognizing the letter A in context, e.g., CAT, than the same letter alone. These results conflict with the naïve assumption that more input should require more processing. This was one of the inconsistencies resulting in the paradigm shift to massively parallel (connectionist) models of brain function.

If there really are fundamental inconsistencies between visual experience (the mind) and the neural theory of the brain, this is a major challenge to the (currently dominant) theories that the mind is constituted entirely from the activity of the brain. As usual, Dennett is unequivocal: "Our minds are just what our brains non-miraculously do "(30, Preface). Not only philosophers are so dogmatic. Stanislas Dehaene, a leading experimentalist, says: "If you had any lingering doubts that your mental life arises entirely from the activity of the brain, these examples should lift them" (37, p.153). No one has suggested how this postulated mind/brain interaction would work, and we will show here that the examples above cannot be explained within existing theories.

*Given an inconsistency between levels of description, it is natural to assume that the deeper (here neural) level is more basic and that the more direct level (here subjective experience, SE) is the problem. Noam Chomsky (43) has effectively challenged this assumption, exploring especially the famous 19<sup>th</sup> century mismatch between Chemistry and Physics, where it required a major (quantum) revolution in Physics to support unification of the two fields. The inconsistency results present here suggest that it might*

*also be the case that solving  the mind-body problem would require major innovation in neurobiology.*

There is always the possibility of a conceptual breakthrough, but it would entail abandoning some of our core beliefs about (at least) neural computation. There is a plausible functional story for the stable world experience and the related binding problem to be discussed below. First of all, we do have an integrated (top-down) sense of the space around us that we cannot currently see, based on memory and other sense data – primarily hearing and smell. In addition, since we are heavily visual, it is adaptive to use vision as broadly as possible. In fact, it would be extremely difficult to act in the world using only the bulls-eye images from Figure 1 and separated information on size, color, etc. The mind (somehow) encodes a more accurate version of the world than can be directly captured by our limited neural hardware.

We should not be surprised that our subjective experience sometimes deviates from the information known to be captured and processed by the visual system. Our senses and the nervous system in general evolved to help our bodies function effectively in a physical and social world that we cannot directly observe (36) . Given that such mental experiences are evolutionarily fundamental, what can we know about their physical realization? There is an impressive body of work suggesting how aspects of subjective experience, closely related to the discussion here, seem to be needed for animals to deal with external space and to combine various sensory inputs and goals. At least some aspects of these experiences may well be found in cephalopods (38) and insects (36). *A detailed discussion of evolutionary considerations can be found in arXiv* http://arxiv.org/abs/2008.08073 *, On the evolution of Subjective Experience.*

This ecological requirement also reveals a deep problem in the use of the term "illusion". The word is sometimes used to describe a perception that is inconsistent with external reality and sometimes to describe an experience that is inconsistent with the neural representation even when the subjective experience is more like external reality. To further confuse the issue, "illusion" is also used metaphorically, as in the postulated "illusion of Free Will" (31). Everyone agrees that we all act as if we had Free Will, even determinists who deny that they have this power. There does not seem to be an agreed definition of "illusion" that supports its use in a serious discussion of the mind.

Computational Limitations

We will now prove that some everyday visual experiences cannot be explained within existing neuroscience. The basic form of the argument will be computational. There is no way that brain neurons, as we know them, could represent or compute the substrate of our visual experience. The constraint of explaining visual experience also rules out many proposed and speculative theories of neural computation in the human brain, as discussed below. To explore the details, we turn next to Figure 4 A, B.

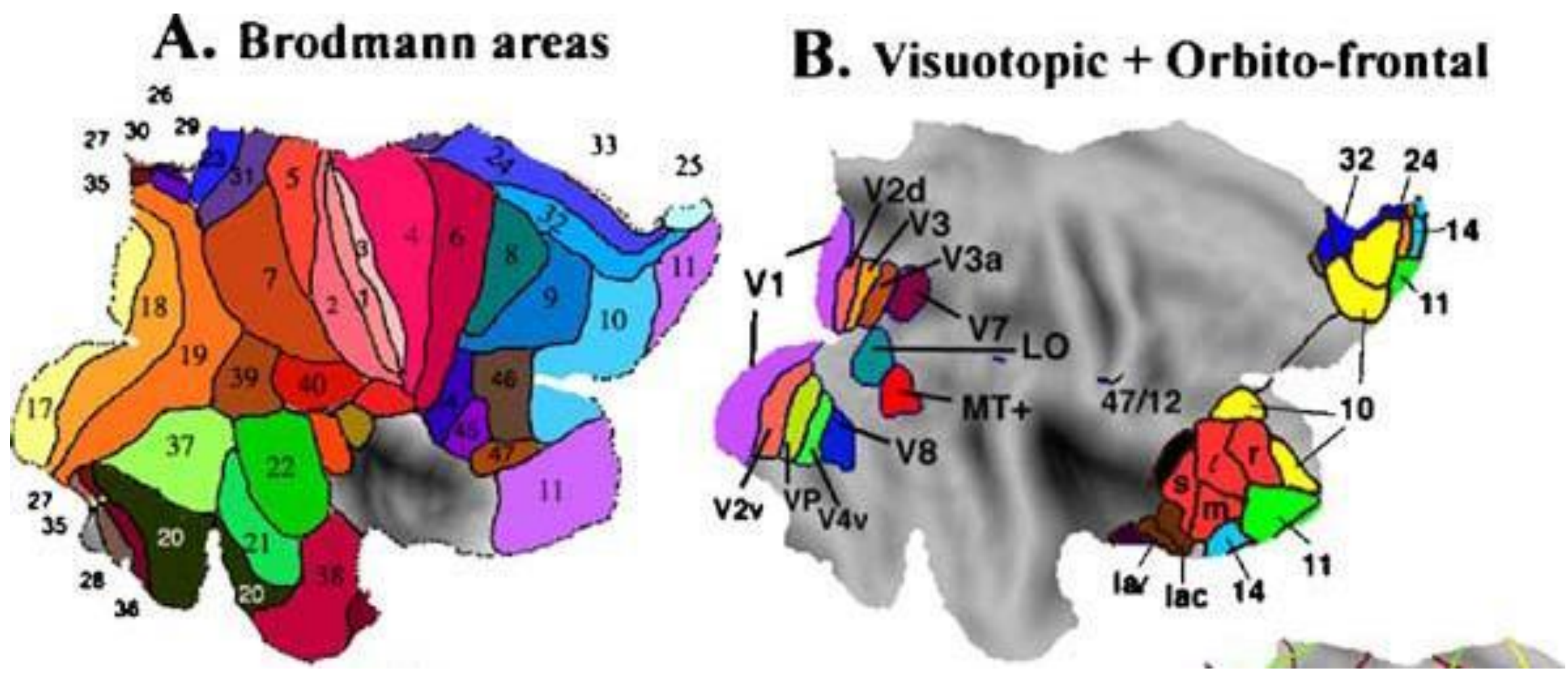

Figure 4 Flat map projection of the Human brain (5)

Figure 4A is a standard flattened projection of one hemisphere of the human brain with the various areas colored. The numbers refer to the traditional Brodmann classification of brain regions from their anatomical details. Modern methods (27, 35) have further refined this picture and elaborated the basic functions computed in these different areas. Figure 4B provides more detail on this functional separation in the visual system, which is at the core of the neural binding problem, one of our mysteries.
The visual area V1, our main concern for the stable world experience and the subject of Figure 3, is shown as the yellow area on the left of Figure 4A (as area 17) and as the magenta area in Figure 4B. Notice that V1 is the largest of the visual areas; this will be important for our discussion.

There are two additional lessons to be gleaned from Figure 4 above. First, 4A shows that the functionality of the cerebral cortex is basically known (5, 27, 35) – there is no large available space for neural computation of currently mysterious phenomena. In addition, various aspects of our visual experience are primarily computed in distinct and often distant interacting circuits. For example, in Figure 4B color calculation is based in the bright green area V4v and motion calculation involves several areas: V3, V3A, MT, etc. In spite of this extreme separation of function, we experience the world as an integrated image with objects that combine all visual properties and even associate these with other senses like sound when appropriate. The mystery of how this happens is called the “hard binding problem” (3).

There are two immediate challenges to address in “the experience of the stable visual world”: apparent stability over saccades and the detailed perception of the full visual field. One popular idea is to suppose that the perceived full field is pieced together as a mosaic of “bull’s-eye” views (Figure 3) from many saccades. There are two serious flaws in this story, one temporal and one spatial, as an explanation of the experience. We only make about 3~4 saccades per second – this is too slow for stable vision (movies are ~ 20 frames per second). In addition, 3 or 4 such images would not yield nearly enough detailed information to build a detailed full field view.

In addition, it would require a huge area of visual neurons to encode the detailed full field view that we subjectively perceive. We can give a quantitative estimate of what is involved. There are a number of alternative calculations, but they all confirm the basic point that fine resolution over a large visual field would require brain area several times larger than V1 (Figure 4).

Stan Klein, who has looked extensively at this issue, suggests the following analysis focusing on the retinal ganglion cells –RGC. The key equations from (14) are:
$Thr(Ecc) = Thr(fovea)*(1 + Ecc/E2)$ or $Sep(Ecc) = Sep(fovea)*(1 + Ecc/E2)$
where Thr (or Sep) is threshold (or separation) in minutes. and Ecc is eccentricity in deg and E2 is the eccentricity at which Thr or Sep double. E2 is the number of degrees of eccentricity at which the spacing of V1 neurons or ganglion cells double. Levi et al. (14) found E2=0.7 deg for cortical cells and is about 1.0 deg for ganglion cells. That is, for ganglion cells the spacing would be $s = 0.5 (Ecc + 1)$ min so at Ecc=0 the spacing is about 0.5 min and at 20 deg it is about 10 min, 20 times as much. This calculation suggests that it might require 20 times as much V1 area to capture the precision of the fovea out to 20 deg of visual angle.

From a slightly different perspective, the cortical magnification factor says that the resolution at 20 degrees eccentricity is 20 times worse than at the foveal projection. This is because of retinal under-sampling in the periphery; the detailed information is only captured at the fovea of the retina (13). Also, the dense neural circuits in the V1 foveal projection has about 200,000 cells per square mm, while at 20 degrees out it is more like 4,000 cells per square mm (15). This is a factor of 50:1 denser in the V1 foveal projection than in the periphery. The V1 foveal projection occupies about a quarter of region on the left in Figure 4. For the brain to encode our detailed perception out to 20 degrees would require an area roughly 12 times the size of V1. There is no way that an area nearly this large could fit into Figure 4. The remarkable recent advances (27, 35) describing a much more detailed parcellation of human cerebral cortex provides even stronger evidence against unknown visual areas.
We can also consider the evidence from the hundreds of full-brain scanning experiments that are exploring which brain locations are active for various vision tasks (16, 17, 18, 35). This precludes the possibility that a network large enough to capture a detailed image could remain undetected.

In summary, as long as we believe that more detail requires more neurons, there is no place in the brain that could encode a basis for the detailed large field image that we experience. This analysis disproves more than the idea of unknown brain circuitry that underlies our stable world experience. It also refutes any plausible substrate for other proposals such as complete “remapping” which suggest that all of the information from one saccade is (somehow) mapped to the input coming from the next saccade (7, p.557).

*However, there is overwhelming experimental and experiential evidence that we do act as if we have a stable detailed internal perceptual model. There is continuing work on*

*looking for some neural realization of this functionality. Early vision is dominantly retinocentric, but the search for explicitly spatiocentric higher visual areas has not been successful. In fact, the organization of higher visual circuits appears to be task oriented (44).*

*" Results from this experiment did not yield retinotopic priming but showed robust spatiotopic and object-centered priming. These findings demonstrate that spatial priming operates within ecologically relevant coordinate systems, and the findings support the notion that spatial priming serves an adaptive role in human behavior."*

*If anything, this increases the mystery of why we have the SE of the stable world.*

The binding problem (3, 41) is a closely related mystery of vision that we can consider, also based on Figure 4. Although the full computational story is more complex, it is the case that different visual features are largely computed in separate brain areas. However, we experience the world as coherent entities combining various properties such as size, shape, color, texture, motion, etc. (39). Again, there is no place in the brain that could encode a detailed substrate for what we effortlessly perceive. This also suggests that our subjective perception (somehow) integrates activity from different brain circuits. Various forms of the binding problem are also the subject of ongoing research (3, 19).

## A Touchstone for alternative brain theories

The discussion above is based on the standard theory (20) that information processing in the brain is based on complex networks of neurons that communicate over long distances mainly by electrical spikes and learn mainly through changes in the connections (synapses) between neurons. This theory also includes a wide range of other chemical and developmental factors, but none that would affect the basic results above.

However, there are a number of alternative proposals that deny the centrality of standard neural computation and several of these are being actively discussed (21, 22, 23); two good sources for a wide range of alternative models are the Journal of Consciousness Studies and http://consciousness.arizona.edu. One reason for this interest in alternative theories is that everyone agrees that the current standard theory does not support a reductionist explanation of historic mind-brain problems like subjective experience and consciousness. The standard theory continues to yield scientific and clinical progress, so any new proposal should be consistent with it. Alternative ideas on the basis for brain information processing include quantum effects (28) and central roles for the glia, for the neuropil, or for microtubules. All of the suggestions for some sub-neural substrate for perception suffer from the same problem – the only known mechanism for the requisite fast long-distance communication in the brain is neural spikes.

Edwards (25) suggests another approach – unified perception (and consciousness) is based on wave patterns in the membranes of individual cells. All of these ideas presuppose that there is some substrate (NCC) for subjective experience in the brain, but there are also more radical theories like that of Alva Noe (22). He claims to explain how "we enjoy an experience of worldly detail that is not present in our brains". Starting from a standard argument that utilities are the basis for perception, Hoffman (33) suggests that: "Your apple and my apple are distinct, just as your headache is distinct from my headache. Something in the objective world triggers us both to perceive an apple, but whatever that thing might be in the objective world, it is almost certainly nonspatial, nontemporal, and in no way resembles an apple".

More general architectural suggestions include global workspace model (40) and the Tononi information model (21). Many proposals suggest some unspecified mass action of neural assemblies, following a long tradition (23). Cohen et al. (26) show how known results on summary statistics and peripheral vision explain some of people's ability to get the "gist" of a scene without capturing all of the detail of subjective perception. After extensive analysis and modeling of peripheral vision and what she calls "the awareness puzzle", Rosenholtz (42) concluded that "Perception is inherently something of an illusion", *but she subsequently (45) rejected any illusion explanation of the puzzle.*

*Recent work examines such proposals. Lamme (46) presents an excellent overview as of 2021 of the main architectural proposals and their prospects for explaining subjective experience and consciousness. Everyone agrees that there are aspects of vision (such as adaptation) of which people are not aware. Based on detailed experimental results, Lamme proposes a 4-level model of visual perception, Recurrent Processing Theory (RPT) (46).*

*" Four stages of vision are identified, going from the fully invisible, to subjectively invisible, unattended, and clearly visible. The relevant feature is a postulated partition of vision into unconscious and conscious perception. It is proposed that feature extraction, categorization, and some aspects of visual inference occur during full and subjective invisibility. It is argued that perceptual organization is the function that is central to understanding the transition from unconscious to conscious seeing. "*

*The article then considers three of the most popular models for the Neural Correlates of Consciousness (NCC). These are: Higher Order Theory , HOTT, (47), Global Network Workspace Theory, GNWT (48), and integrated Information Theory, IIT, (49). Lamme shows that none of these models explicitly specifies the partition between aware and unaware experience that is needed for his analysis. By design, Lamme's own RPT does identify "perceptual organization" as the key to consciousness. This does not solve the mystery of visual experience, but is a necessary step. Independent of Lamme's particular model, we should expect proposed NCC solutions to specify the basic awareness partition.*

Since the deep mind-brain phenomena of most common interest are not well defined, it has not been straightforward to evaluate any of these suggested alternatives to the standard model of neural computation. *There is increasing interest in the history and current activity in theories and models of the mind. An excellent overview is available as "The Idea of the Brain: A History" (54)*

The findings described in early sections above could yield concrete touchstone problems for proposed theories of representation, computation, and communication in the brain. Both the binding problem and the experience of a detailed stable visual scene are ubiquitous in daily life, are functionally necessary, and have clear informational requirements. We could ask proponents of speculative brain models how their theory could account for these two concrete phenomena. That is, assume your theory is true and show how it helps explain these (or other) touchstone problems. I have done this informally with several leading proponents of alternative models and have never heard even a vague claim of adequacy.

The general acceptance of some such touchstone tasks could sharpen the discussion of information processing in the brain. Of course, the deep mind-brain problem remains a mystery, but we should expect proposed models of neural computation to address some of the concrete touchstone problems, like those discussed here.

Experiments

We have shown that existing neuroscience cannot address all the basic questions of subjective experience. Of course, very productive communities are working on many aspects of cognitive and perceptual neuroscience. There is a wide range of motivations for these efforts, but these do not usually include trying to elucidate remaining mysteries, such as the ones discussed here. There are risks involved in attacking difficult problems and there are relatively few such efforts in the current stressful research environment.

*The longstanding mystery of "filling out" has been examined in a recent experiment (51). "Participants perceived that peripheral stimuli changed to match the central stimuli, so that the display seemed uniform… a wide range of visual features, including shape, orientation, motion, luminance, pattern, and identity, are susceptible to this uniformity illusion" They suggest that the uniformity illusion is the result of a reconstruction of sparse visual information (from the periphery) based on more readily available detailed visual information (from the fovea), which gives rise to a rich, but illusory, experience of peripheral vision. Of course, this is another mystery of visual experience.*

Nevertheless, there is some research that is making progress on demystifying some of the mysteries of subjective experience. In the Fall of 2017, a UC Berkeley interdisciplinary seminar course explored "Science and Subjectivity " http://rctn.org/wiki/VS298:_Subjectivity, *still available in2021* . This web site for the course contains references and video lecture recordings for a wide range of research that pushes on the boundary between routine science and remaining mysteries. Week 3 contains lecture and discussion on the material in this article. The web site also includes presentations and discussion by Michael Cohen, Jerry Feldman, Rich Ivry, Stan Klein,

Bob Knight, Christof Koch, Ken Nakayama, Brian Odegaard, Bruno Olshausen, Terry Regier, Shin Shimojo, and Peter Tse. If there is a scientific explanation of these mysteries, it will need to be an evolutionary story, so it should be fruitful to focus on a wide range of animals (36, 38). *This is discussed in On the Evolution of Subjective Experience* http://arxiv.org/abs/2008.08073 *.*

*The (2016) first version of this article focused on the whole visual system and showed that the SE of a full detailed scene is incompatible with current cognitive neuroscience. However, it did not explicitly address the various levels of processing or the relation between foveal and peripheral vision. New work on the processing levels by Lamme (46) is discussed earlier in this version. The foveal-peripheral issue has been addressed in an elegant set of experiments and computational studies (45). The important issue here is whether the two visual systems and visual memory can suggest solutions to the mysteries explored in this paper.*

*An excellent discussion of peripheral vision is "A review of interactions between peripheral and foveal vision." ( 45). The article demonstrates that these interactions are bidirectional and ubiquitous. Their paper considers several issues that bear on our mysteries. Peripheral vision captures the majority of the visual field, but only coarsely. "Overall, these results illustrate that peripheral processing is only sufficient for coarse analysis and recognition, and that foveal processing is necessary for the analysis of object details. " A core question is how the coarse and fine results are combined, including over saccades. "The weighting with which pre- and post-saccadic information is combined is not, however, equal, rather it occurs in proportion to the relative reliability of each stimulus."*

*Despite all the additional experiments and analysis. Stewart et al. have no constructive suggestion on the SE of our core mystery. "How this homogeneity of visual experience comes about is still debated intensely. One possible interpretation is that perception in the periphery merely appears sharper"*

*Rosenholz (50) presents a compatible detailed computational analysis of peripheral vision and suggests a novel approach to the mystery of SE. The first phase of her approach is based on the considerable literature on the peripheral perception of "gist" of complex images (50). The new idea is based on postulating a post-perceptual cognitive decision phase that is also subject to overload.*

*Rosenholtz uses this 2-phase model to help explain some specific experimental findings. However, it does not directly address the core SE mysteries.*

*So, continuing extensive analysis of structural and functional analysis of the visual system, including the work of Rosenholtz and of Lamme(46) discussed above, has not yet yielded further insights into the core mysteries.*

Conclusions

There is general agreement that there are mysteries about the world and our place in it that are not yet understood. Even radical materialists will concede that there are questions (e.g., free will) that might never have scientific solutions. Nevertheless, it is not widely understood that, every time we open our eyes, we experience phenomena that cannot be explained with existing neuroscience and possibly not with any science.

As thinkers, we have no choice but to acknowledge that we do not know and may never know the answers to many deep questions about the world and ourselves (24). There are two basic ways to learn about the physical and social world: investigation and stories. Science is a uniquely powerful tool of investigation, but is limited in scope at least at present. Of course, there remains a vast amount that can and should be explored and exploited scientifically. Stories can provide insights that are not directly testable, and this is certainly also important in science. The stories in art, mythology, religion, etc. have been and will remain powerful sources of guidance about how to live.

Initial attempts to convey the ideas in this note have not been very successful. I find that scientists have a strong negative reaction to my simple demonstrations. Philosophers and humanists welcome any attack on reductionism but prefer argument to science. Both responses seem mainly territorial. It does not appear to be too much of a stretch to view the (Eastern or Western) religious practitioners as other special interest groups. It is certainly true that all these groups have emotional/spiritual as well as financial/power stakes in the big questions, and there does not seem to be much support for the agnostic mysterian position "we simply don't know".

*From my 2021 perspective, it is useful to consider Subjective Experience(SE) from the perspective of Tinbergen's famous four criteria for traits ( https://en.wikipedia.org/wiki/Tinbergen%2s_four_questions) (1) adaptive function, (2) phylogenetic history, (3) underlying physiological mechanisms and (4) ontogenetic/developmental history. There are many other cases of SE where your experience inconsistent with current neuroscience. For example, if you touch your nose with your thumb, it is experienced as simultaneous despite the large difference in neural conduction times ( also cf. 53). The adaptive function (1) of such SE is to provide a useful internal substrate for action and planning. The evolutionary(2) and developmental (4) aspects of SE are discussed in detail in On the Evolution of Subjective Experience* http://arxiv.org/abs/2008.08073*. It is only the realization (3) that remains mysterious.*

Everyone is entitled to his or her own (religious or other) beliefs, but there is nothing in our current ignorance that privileges one faith over all others. Belief in the inevitability of complete scientific answers (8, 21, 30) is one such faith. There are beliefs (e.g. about the age of the Earth) that contradict established scientific knowledge and cannot be taken seriously. Somewhat surprisingly, there seems to be little support in cognitive and neuro science for "we simply don't know". This is despite the fact "we simply don't know and may never know" is the accepted response in physics for some fundamental questions. What has been most surprising is how many people prefer believing there must be an (inscrutable) reductionist answer rather than accepting the agnostic stance.

Even philosophers, who consider the mind/brain problem as a possible feature of nature, try to prove their contention rather than leave it as one possibility (8). This all might be related to recent findings suggesting an innate human drive to find answers for unanswerable questions (32).

Nevertheless, in the face of all that is unknown, we (following the physicists) would do well to appreciate both what is scientifically known and also the mysteries that remain. Ideally, results like those above will encourage theory and experiment on questions at the boundaries between the known and unknown. *The most promising way forward is the continuing detailed experiment and modeling of currently mysterious phenomena, as exemplified by " The macaque face patch system: a turtle's underbelly for the brain" (52).*

Acknowledgement
This work was supported in part by the Office of Naval Research under Grant N000141110416 and by a grant from Google.